\documentclass[aps,prb,superscriptaddress,twocolumn,showpacs,amsmath,amssymb]{revtex4}

\usepackage{amsmath}
\usepackage{color}
\usepackage{graphicx}
\usepackage{bm}
\definecolor{Blue}{rgb}{0.00, 0.00, 1.00}
\definecolor{Red}{rgb}{1.00, 0.00, 0.00}

\begin{document}

\title{Topological currents in black phosphorus with broken inversion symmetry}

\author{Tony Low}
\affiliation{Department of Electrical \& Computer Engineering, University of Minnesota, Minneapolis, MN 55455, USA}
\author{ Yongjin Jiang}
\affiliation{Department of Electrical \& Computer Engineering, University of Minnesota, Minneapolis, MN 55455, USA}
\affiliation{Center for Statistical and Theoretical Condensed Matter Physics, ZheJiang Normal University, Jinhua 321004, Peoples Republic of China}
\author{ Francisco Guinea}
\affiliation{School of Physics \& Astronomy, University of Manchester, Oxford Road, Manchester, M13 9PL, UK}
\affiliation{IMDEA Nanociencia, Faraday 3, 28049 Madrid, Spain}

\date{\today}

\begin{abstract}
We examine the nature of topological currents in black phosphorus when its inversion symmetry is deliberately broken. Here, the conduction and valence band edges are located at the $\Gamma$ point of the rectangular Brillouin zone, and they exhibit strong anisotropy along its two crystal axes. We will show below that these salient features lead to a linear transverse neutral topological currents, accompanied also by a  non-linear transverse charge current at the Fermi surface. These topological currents are maximal  when the in-plane electric field is applied along the zigzag crystal axes, but zero along the armchair direction.

\end{abstract}


\maketitle

Topological currents is one of the well-known physical manifestation when a crystalline solid possesses a finite Berry curvature\cite{berry1984quantal,xiao2010berry,nagaosa2010anomalous}. The Berry curvature is a geometrical property of the Bloch energy band, and acts as an effective  magnetic field in momentum space\cite{berry1984quantal}. Hence, topological materials may exhibit anomalous Hall-like transverse currents in the  presence of an applied electric field, in absence of a magnetic field. In topological insulators\cite{hasan2010colloquium}, topological bands with non-trivial Berry phase leads to propagating surface states that are protected against backscattering from disorder and impurities. In transition metal dichalcogenides, the two valleys carry opposite Berry curvature, or magnetic moment, giving rise to a bulk topological charge neutral valley currents\cite{xiao2012coupled,mak2014valley,lensky2014topological}. Recently, these bulk topological currents were also experimentally investigated in other Dirac materials, such as gapped graphene and bilayer graphene system\cite{gorbachev2014detecting,sui2015gate}.

In this letter, we examine the nature of the topological currents in black phosphorus (BP)\cite{morita1986semiconducting,li2014black,liu2014phosphorene}, when its inversion symmetry is deliberately broken. Unlike more well-studied 2D materials such as graphene and transition metal dichalcogenides, several key differences are notable. First, in monolayer BP, the conduction and valence band edges are located at the $\Gamma$ point of the rectangular Brillouin zone\cite{rudenko14}. Second, the energy bands  exhibit strong anisotropy along its two crystal axes. We will show below that these salient features will lead to a linear transverse \emph{neutral} topological currents, accompanied also by a non-linear transverse \emph{charge} current. These
topological currents are maximal (zero) when the in-plane electric field is applied along the zigzag (armchair) crystal axes. We discuss how these topological effects can be detected electrically and optically.

\begin{figure}[t]
   \centering
	\scalebox{0.65}[0.65]{\includegraphics*[viewport=215 190 580 470]
	{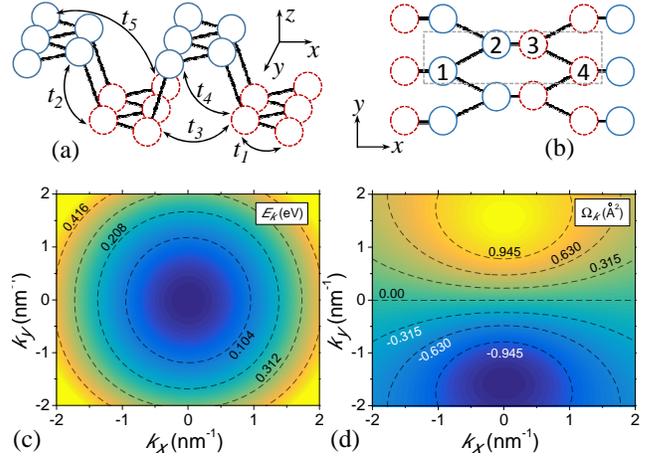}}
 \caption{(a) Crystal structure of monolayer black phosphorus, and the hopping parameters included in the tight-binding model. Here, $x$ is along the so-called armchair direction, while $y$ is along zigzag. (b) Top view of the crystal structure and the primitive unit cell is also indicated.
(c) and (d) plots the energy dispersion, $E_{\bold{k}}$, and its Berry curvature, $\Omega_{\bold{k}}$, for the conduction band in the vicinity of the band minimum at $\Gamma$ point.
Inversion symmetry is broken by applying the following potentials to the basis atoms, $V_{1,4}=0\,eV$, $V_{2}=-0.5\,eV$, and $V_{3}=0.5\,eV$.
 }
 \label{fig1}
\end{figure}

In this work, we consider monolayer BP, described with a four-band tight-binding model\cite{rudenko14} as illustrated in Fig.\,1a.
We construct Bloch-like basis functions $\left|\chi_j^\bold{k}\right\rangle = \sum_{\bold{R}} \mbox{exp}[i\bold{k}\cdot(\bold{R}+\bold{b}_j)]  \left|\phi_j^{\bold{R}}\right\rangle$
where $\bold{R}$ and $\bold{b}_j$ are the lattice and basis vectors, and
the index $j$ runs over all the phosphorus atoms within the primitive cell,
with a single orbital on each atom denoted by $\left|\phi_j^{\bold{R}}\right\rangle$ . The Hamiltonian matrix can then be constructed from,
\begin{eqnarray}
{\cal H}_{ij}(\bold{k})=\left\langle \chi_j^\bold{k}\right| {\cal H} \left|\chi_j^\bold{k}\right\rangle=\sum_{\bold{R}} e^{i\bold{k}\cdot(\bold{R}+\bold{b}_j-\bold{b}_i)}h_{ij}
\end{eqnarray}
where $h_{ij}\equiv \left\langle \phi_i^{\bold{0}}\right| {\cal H} \left|\phi_j^{\bold{R}}\right\rangle $ contains the tight-binding hopping parameter.   The hopping parameters\cite{rudenko14} used in this work are $t_1=-1.22\,eV$, $t_2=-3.67\,eV$, $t_3=-0.205\,eV$, $t_4=-0.105\,eV$, and $t_5=-0.055\,eV$. The secular equation to be solved is ${\cal H}_{\bold{k}} \left|\Psi_{n\bold{k}}\right\rangle =  E_{n\bold{k}} \left| \Psi_{n\bold{k}}\right\rangle $, where ${\cal H}_{\bold{k}}$ is the $4\times 4$ matrix of elements ${\cal H}_{ij}(\bold{k})$, $\left|\Psi_{n\bold{k}}\right\rangle$ and $E_{n\bold{k}}$ are the eigenvectors and eigen-energies.

The Berry curvature for the electronic Bloch states of the $n^{th}$ band can then be computed from\cite{xiao2010berry},
\begin{eqnarray}
\bold{\Omega}_n(\bold{k})=\Omega_n(\bold{k})\hat{\bold{z}}=\nabla_{\bold{k}}\times \left\langle \Psi_{n\bold{k}}\right|i \nabla_{\bold{k}} \left|\Psi_{n\bold{k}}\right\rangle
\end{eqnarray}
and it's magnitude has the following explicit form,
\begin{eqnarray}
\Omega_n(\bold{k})=2\Im\!\!\left[\sum_{m\neq n} \frac{\left\langle \Psi_{n\bold{k}}\right|i \partial_{k_x}\left|\Psi_{m\bold{k}}\right\rangle  \left\langle \Psi_{m\bold{k}}\right|i \partial_{k_y}\left|\Psi_{n\bold{k}}\right\rangle }{(E_{m\bold{k}}- E_{n\bold{k}})^2}\right]
\end{eqnarray}
Time reversal symmetry implies that $\bold{\Omega}_n(\bold{k}) = -\bold{\Omega}_n(-\bold{k})$, while crystal lattice with inversion symmetry would requires $\bold{\Omega}_n(\bold{k}) = \bold{\Omega}_n(-\bold{k})=0$. Hence, inversion symmetry breaking is necessary to generate a finite Berry curvature.

We consider some basic symmetry properties of the Hamiltonian. When the on-site potentials $V_j$ are zero, there are two inversion centers i.e. between atom $1$ and $2$, and atom $2$ and $3$. We denote these space inversion symmetries as ${\cal P}_1$ and ${\cal P}_2$ respectively.
In addition, we have mirror symmetry operations ${\cal M}_x$, which interchange atoms $1$ and $2$ with atoms $4$ and $3$ respectively, and ${\cal M}_y$ which is diagonal in the atomic species space. We constraint ourselves to inversion symmetry breaking schemes with only electrostatic on-site potentials $V_j$ within each unit cell. First, consider a perpendicular electric field, i.e. $V_1=V_2\neq V_3=V_4$, it breaks ${\cal M}_x$ and ${\cal P}_2$, but not ${\cal M}_y$ and ${\cal P}_1$. Second, consider electrostatic potentials staggering along $y$ (zigzag) direction, i.e. $V_2=V_3\neq V_1=V_4$. This scheme breaks only ${\cal P}_1$, but not ${\cal P}_2$, ${\cal M}_x$ and ${\cal M}_y$. On the other hand, an electrostatic potential staggering along $x$ (armchair) direction will break all inversion symmetries, and ${\cal M}_x$, except for ${\cal M}_y$. Although this configuration can generate non-zero Berry curvature, additional symmetries ensure zero transverse currents in this case, a subtle point we will elaborate below.


Alternatively, the combination of an out-of-plane electric field and an in-plane lattice commensurate electric field directed along $y$ direction can break all inversion symmetries, since $V_4-V_3=V_1-V_2\neq 0$. For example, let's consider  $V_{1,4}=0\,eV$, $V_{2}=-0.5\,eV$, and $V_{3}=0.5\,eV$. Fig.\,1c-d plots the energy dispersion, $E_{\bold{k}}$, and its Berry curvature, $\Omega_{\bold{k}}$, for the conduction band in the vicinity of the band minimum at $\Gamma$ point. The smaller electron effective mass along the armchair direction, $x$, leads to stronger dispersion as shown. Indeed, $\Omega_{\bold{k}}$ is finite, and can take either signs in the Fermi sea of the $\Gamma$ valley. This is reminescent of the conventional valley physics, e.g. in gapped graphene and transition metal dichalcogenides where the Berry curvature of the two valleys bear opposite signs, except we have only single valley in this case.

In the above-mentioned scheme, the mirror symmetry ${\cal M}_y$ is maintained, even when its inversion symmetries are being deliberately broken.
Time reversal symmetry requires that its energy dispersion respects $E_n(\bold{k})= E_n(-\bold{k})$. The above-mentioned mirror symmetry would entails $E_n(k_x,k_y)= E_n(k_x,-k_y)$. These symmetries in combination also implies $E_n(k_x,k_y)= E_n(-k_x,k_y)$. 
For the Berry curvature, which is anolagous to an effective magnetic field, to produce physical observable that respect this ${\cal M}_y$ mirror symmetry would require  $\bold{\Omega}_n(k_y) = -\bold{\Omega}_n(-k_y)$. Time reversal symmetry would then impose the additional constraint that $\bold{\Omega}_n(k_x) = \bold{\Omega}_n(-k_x)$. This accounts for the form of the computed Berry curvature shown in Fig.\,1d.

\begin{figure}[t]
   \centering
	\scalebox{0.8}[0.8]{\includegraphics*[viewport=230 140 580 520]
	{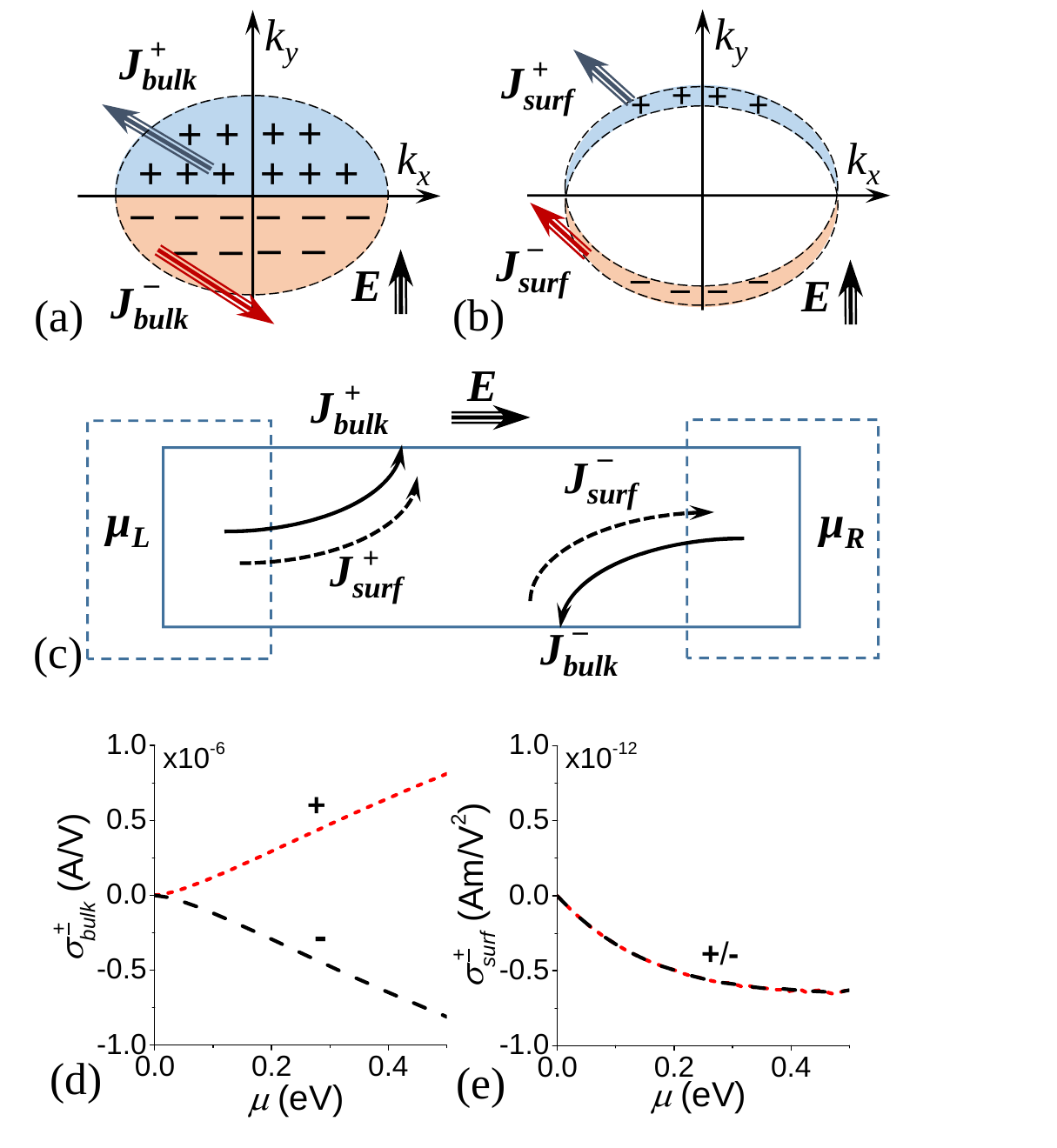}}
 \caption{(a) `Bulk Fermi sea' contribution to the transverse current, which can be partitioned into the forward/backward propagating currents  ${\cal J}_{bulk}^\pm$. These current are linear with electric field $\xi$, and persists even in equilibrium. (b) `Fermi surface' contribution, also be partitioned into ${\cal J}_{sur}^\pm$, has a $\xi^2$ dependence, and is a non-equilibrium phenomenon. (c) Illustration of the various transverse current components in a typical two terminal device. (d) and (e) plots the dependence of the bulk and surface transverse currents with Fermi level $\mu$, expressed in terms of their conductivities (see text). Calculations assumed temperature of $10\,$K.
 }
 \label{fig2}
\end{figure}

Following these considerations, we examine the nature of topological currents arising from the induced Berry curvature,  within the semiclassical Boltzmann transport theory.
In the presence of an external electric field $\boldsymbol{\xi}$, the carrier velocity acquires a non-classical transverse term due to Berry curvature as shown\cite{xiao2010berry},
\begin{eqnarray}
\bold{v}_n(\bold{k})=\frac{1}{\hbar}\bold{\nabla}_{\bold{k}}E_{n\bold{k}}
-\frac{e}{\hbar}\boldsymbol{\xi}\times \bold{\Omega}_n(\bold{k})
\end{eqnarray}
The transverse currents can be partitioned into contributions from the forward and backward propagating states, denoted by ${\cal J}^{\pm}$. By the former (latter), we refer to states whose $\bold{v}_n(\bold{k})$ is such that $S\equiv\mbox{sign}(\bold{v}_n\cdot \boldsymbol{\xi})=\pm$ respectively. ${\cal J}^{\pm}$ can be computed semiclassically up to second order in the electric field, i.e. $\sigma_{bulk}^{\pm} \xi + \sigma_{surf}^{\pm} \xi^2 +O(\xi^3)$, with the conductivities defined as,
\begin{eqnarray}
\nonumber
\sigma_{bulk}^{\pm}= -\frac{e}{\hbar}\int_{S=\pm } d\bold{k} f_0(\bold{k})\Omega(\bold{k})\\
\sigma_{surf}^{\pm}= \frac{e^2\tau}{\hbar^2}\int_{S=\pm } d\bold{k} [\nabla_{\bold{k}}f_0(\bold{k})\cdot \boldsymbol{\xi} ]\Omega(\bold{k})
\end{eqnarray}
where $f_0(\bold{k})$ is the Fermi-Dirac distribution function, and $\tau$ is the electron scattering time. The linear contribution to ${\cal J}$ is a `bulk Fermi sea' phenomenon, and can be partitioned into the forward/backward propagating currents  ${\cal J}_{bulk}^\pm$ as illustrated in Fig.\,2a. These are counter-propagating currents which persist even when the system is in equilibrium. The transverse currents are certainly charge neutral since ${\cal J}_{bulk}^+= -{\cal J}_{bulk}^-$. Nevertheless, recent experiments have shown that such transport effects can be detected via a non-local transport measurement\cite{gorbachev2014detecting,sui2015gate}.

On the other hand, the nonlinear contribution to ${\cal J}$ is a `Fermi surface' contribution, and can also be partitioned into ${\cal J}_{sur}^\pm$, see Fig.\,2b. These currents, however, are non-equilibrium in nature, and requires an electrochemical potential bias. For example, Fig.\,2c illustrates the flows of the various current components in a typical two terminal device under bias. Unlike the bulk currents, the latter has a net charge current since ${\cal J}_{sur}^+= {\cal J}_{sur}^-$. We contrast this with conventional valley physics  where Fermi surface contributions lead to transverse charge neutral currents instead, due to exact cancellation from the two valleys. Recent work found that finite nonlinear current can arise when the two valleys are not isotropic\cite{yu2014nonlinear}, leading to partial cancellation. In a single valley system like BP, the effect will be maximal.

\begin{figure}[t]
   \centering
	\scalebox{0.6}[0.6]{\includegraphics*[viewport=160 150 580 500]
	{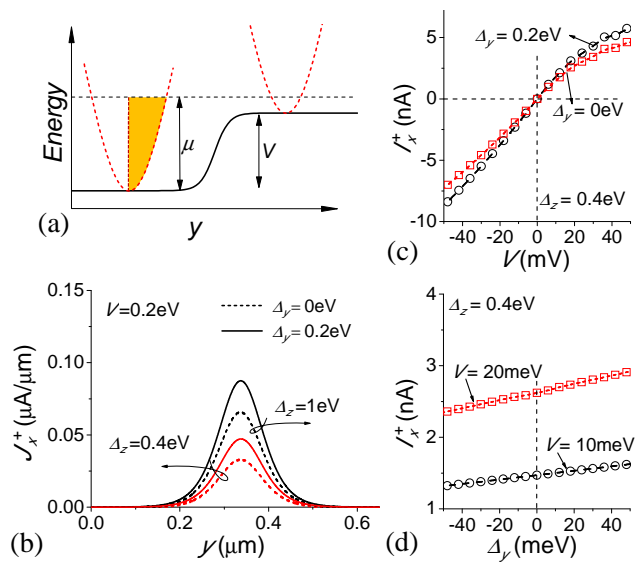}}
\caption{(a) Schematic illustrating the electrostatic junction at equilibrium, along the zigzag $y$ direction. Fermi energy $\mu$ and built-in potential $V$ are indicated. (b) Distribution of transverse current density ${\cal J}^{+}_{xy}$, flowing along $x$, plotted across the junction, under different applied out-of-plane potential $\Delta_z$ and crystal potential $\Delta_y$ as indicated (see text).  (c) and (d) show the integrated transverse current under different applied $V$ and $\Delta_y$ respectively. All calculations assumed $\mu=0.5\,$eV and zero temperature.  }
 \label{fig3}
\end{figure}

The symmetry of the Berry curvature, i.e. $\bold{\Omega}_n(k_y) = -\bold{\Omega}_n(-k_y)$ and $\bold{\Omega}_n(k_x) = \bold{\Omega}_n(-k_x)$, has important consequences on the orientation dependence of the various transverse currents. ${\cal J}_{bulk}^\pm$ and ${\cal J}_{sur}^\pm$ will be zero when the electric field $\boldsymbol{\xi}$ is directed along the armchair direction. These transverse currents attained their maximum when $\boldsymbol{\xi}$ is directed along zigzag direction.

Fig.\,2d-e plots the dependence of these bulk and surface transverse currents with Fermi level $\mu$, expressed in terms of their conductivities. In 2D electron gas, $\mu \propto n$, where $n$ being the carrier density. We found that $\sigma_{bulk}\propto \mu$ and $\sigma_{surf}\propto \sqrt{\mu}$. This is consistent with the fact that the number of bulk and surface Fermi states scales with $\mu$ and $\sqrt{\mu}$ respectively. The computed linear response is an order of magnitude smaller than that predicted for transition metal dichalcogenides\cite{xiao2012coupled}. However, the finite nonlinear response in this case can produce a comparable or larger effect at higher driving fields.

To gain deeper insights into the above-mentioned issues, we consider an edge-free model system, an electrostatic junction with a built-in electric field aligned along the zigzag (i.e. $y$) direction. The electrostatic junction provides a built-in electric field that drives the transverse bulk current response. Here, we consider the device under equilibrium condition.


Our approach solves the quantum mechanical scattering problem microscopically within the above-mentioned tight-binding model. We describe this built-in junction with $V(y)=V\mbox{tanh}(\tfrac{x}{\alpha})$, as shown in Fig.\,3a, where $V$ is the built-in potential. \textcolor{black}{In this calculation, we assumed a junction transition length of $\alpha=60\,$nm.} The Fermi energy, $\mu$, is biased within the lowest conduction band of BP. We consider the combined effect of an out-of-plane electric field, $V_1=V_2=\tfrac{1}{2}\Delta_z$ and $ V_3=V_4=-\tfrac{1}{2}\Delta_z$, and a periodic crystal potential commensurate with the BP unit cell, $V_1=V_4=\tfrac{1}{2}\Delta_y$ and $ V_2=V_3=-\tfrac{1}{2}\Delta_y$.
Periodic boundary condition is imposed along $x$, hence $k_x$ can be regarded as a good quantum number. Assuming semi-infinite perfect leads, then what we have is essentially a one-dimensional quantum scattering problem.  The scattering wave functions of the tight-binding problem, as well as key local observable (e.g. current density, the quantity of interest here), can be solved numerically through standard approaches\cite{di2008electrical,datta1997electronic}.  In the following, we show that a transverse current can indeed flows along the junction, i.e. $x$ direction, when appropriate symmetries are broken in consistent with the semiclassical discussions above.


We are interested in the transverse currents arising from the forward propagating states deep in the Fermi sea, as depicted by the shaded part of the energy dispersion in Fig.\,3a. Here, all calculations assumed $\mu=0.5\,eV$, and zero temperature.
In Fig.\,3b, we plot the calculated transverse current density profile ${\cal J}^{+}_{x}(y)$ across the junction, assuming $V=0.2\,eV$.
When $\Delta_y$ and $\Delta_z$ are non-zero, crystalline inversion symmetry is broken and a finite Berry curvature is present. Indeed, a finite transverse current can be observed, which peaks at the middle of the junction where electric field is maximum. Away from the junction, ${\cal J}^{+}_{x}$ goes to zero.

Symmetry arguments inform us that the transverse current should be odd with respect to $V$ i.e. ${\cal J}^{\pm}_{x}(-V) = -{\cal J}^{\pm}_{x}(V)$. The Hamiltonian describing the device with the built-in junction is related by the mirror symmetry ${\cal M}_y$, in the small $V$ limit. Since  ${\cal M}_y$ does not affect the transverse current, we have ${\cal J}^{+}_{x}(-V) = {\cal J}^{-}_{x}(V)$. On the other hand, time reversal symmetry would require ${\cal J}^{+}_{x}(V) = -{\cal J}^{-}_{x}(V)$. Taken together, we have ${\cal J}^{+}_{x}(-V) = -{\cal J}^{+}_{x}(V)$. Hence, the response to a small applied $V$ should be linear. Numerical results shown in Fig.\,3c confirms this simple argument. 

\begin{figure}[t]
   \centering
	\scalebox{0.65}[0.65]{\includegraphics*[viewport=210 300 580 470]
	{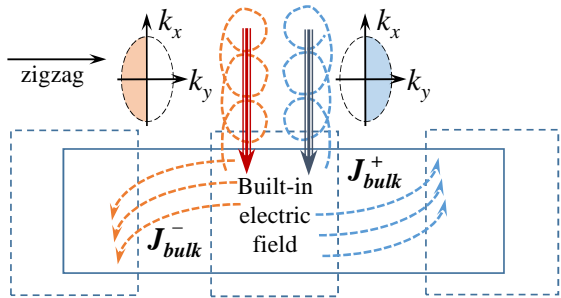}}
\caption{ Schematic illustrating the excitations of states with positive or negative Berry curvatures with circularly polarized light. For example, right circularly polarized light couples to the states with positive Berry curvatures, which will produce an electrical current flowing towards the right contact and the top edge. The transversal currents changes sign with $\Delta_z$.  }
 \label{fig4}
\end{figure}

From the numerics, we found that the transverse current have two distinct contributions, i.e. ${\cal I}^{+}_{x}\sim   \Delta_z\xi  +  \Delta_y\Delta_z\xi$. These trends can be observed in Fig.\,3c and d.
The latter contribution is analogous to the `bulk Fermi sea' semiclassical ${\cal J}_{bulk}$ we discussed earlier\cite{note}. At small $V$, we clearly distinguish the linear regime consistent with the semiclassical result ${\cal J}_{bulk}\propto \xi$. \textcolor{black}{We observed a rollover in ${\cal I}^{+}_{x}$ at larger positive $V$, as more states deep in the Fermi sea are completely reflected due to the energy barrier.} When $V$ is negative, there is no energy barrier, hence the linear trend persists. The former contribution to ${\cal I}^{+}_{x}$ 
is a residual transverse current that probably has a different origin. A finite $\Delta_z$ breaks ${\cal M}_x$ symmetry and produces a current along $x$. Since the magnitude of ${\cal I}^{+}_{x}$ is tunable by varying $\Delta_z$, it provides an obvious way for the detection of the proposed effect.
The proposed effect can also be detected optoelectronically. Fig. 4 illustrates a possible experimental scheme. Circularly polarized light can couples preferentially to states with Berry curvatures of particular sign, and producing a longitudinal and transverse electrical current.


Last but not least, we discuss some considerations on the experimental observation of this effect. Breaking of the crystal inversion symmetry is key. Finding the appropriate substrate which is commensurate along the zigzag direction of BP is needed to provide a finite $\Delta_y$.
Certainly, high mobility samples  are desirable for the observation of the proposed effect.
Recently, encapsulation of BP with hexagonal boron nitride\cite{cao15}, all within a controlled inert atmosphere, has allowed for higher carrier mobilities\cite{exp1,exp2,exp3,exp4,cao15}. Indeed, high quality BP has made possible the first observation of prominent quantum magneto-oscillations in these devices\cite{exp1,exp2,exp3,exp4,cao15}. The results we obtained here are also applicable to other emerging 2D materials with broken inversion symmetry and anisotropic bands at the $\Gamma$ valley\cite{tongay2014monolayer,island2015tis3}.

\emph{Acknowledgement.} TL and YJ acknowledge support from the University of Minnesota start-up fund. FG acknowledges support from the Spanish Ministry of Economy (MINECO) through Grant No. FIS2011-23713, the European Research Council Advanced Grant (contract 290846), and the European Commission under the Graphene Flagship, contract CNECTICT-604391.


\begin{thebibliography}{24}
\bibitem[{not()}]{note}
\bibinfo{note}{It is worth noting that $\Delta_z$ suffices to induce transverse
  currents. This term breaks geometrical spatial inversion, but not the
  effective inversion symmetry of the Hamiltonian discussed in the text.}
\end{thebibliography}

\end{document}